\documentclass[11pt]{article}
\usepackage{graphicx}
\usepackage{amssymb}
\usepackage{amsmath, amsthm}
\newcommand{\be}{\begin{equation}}
\newcommand{\ee}{\end{equation}}
\newcommand{\word}[1]{\,\,\mbox{#1}\,\,}
\newcommand{\reff}[1]{(\ref{#1})}
\newcommand{\beq}{\begin{equation}}
\newcommand{\eeq}[1]{\label{#1}\end{equation}}
\newcommand{\beg}{\begin{equation*}}
\newcommand{\eeg}{\end{equation*}}

\newcommand{\less}{\!<\!}

\newcommand{\sumprime}{\sideset{}{'}\sum}

\newcommand{\bsplit}{\begin{split}}
\newcommand{\esplit}{\end{split}}

\begin{document}
\def\theequation{\arabic{section}.\arabic{equation}}
\begin{titlepage}
\title{Casimir piston for massless scalar fields \\in three dimensions}
\author{\thanks{Email: aedery@ubishops.ca} Ariel Edery \\
{\small \it Physics Department, Bishop's University}\\
{\small \it 2600 College Street, Sherbrooke, Qu\'{e}bec, Canada
J1M~0C8}}

\date{} \maketitle
\thispagestyle{empty} \vspace*{1truecm}

\begin{abstract}
\noindent We study the Casimir piston for massless scalar fields
obeying Dirichlet boundary conditions in a three dimensional cavity
with sides of arbitrary lengths $a,b$ and $c$ where $a$ is the plate
separation. We obtain an exact expression for the Casimir force on
the piston valid for any values of the three lengths. As in the
electromagnetic case with perfect conductor conditions, we find that
the Casimir force is negative (attractive) regardless of the values
of $a$, $b$ and $c$. Though cases exist where the interior
contributes a positive (repulsive) Casimir force, the total Casimir
force on the piston is negative when the exterior contribution is
included. We also obtain an alternative expression for the Casimir
force that is useful computationally when the plate separation $a$
is large.
\end{abstract}
\vspace*{1truecm} %
\noindent
\begin{center}
%{\bf PACS:} 98.62.Sb, 04.90.+e
% Grav. lenses and luminous arcs
% Other topics in GR and gravitation
\end{center}
%\begin{center}  {\bf Keywords:}
%\end{center}
\setcounter{page}{1}
\end{titlepage}

\def\theequation{\arabic{section}.\arabic{equation}}

%%%%%%%%%%%%%%%%%%%%%%%%%%%%%%%%%%%%%%%%%%%%%%%%%%%%%%%%%%%%%%%%%%%%%%%%%%%%%%%%%%%%

\setcounter{page}{2}

\section{Introduction}
\setcounter{equation}{0}

Two years ago, in an interesting paper \cite{Cavalcanti}, the
Casimir piston was studied for a two-dimensional scalar field
obeying Dirichlet boundary conditions on a rectangular region. Among
other things, it was shown that the Casimir force on the piston is
always attractive (negative) regardless of the ratio of the two
sides. In this paper, we study the three-dimensional Casimir piston
for massless scalar fields. A Casimir piston in three dimensions is
depicted in Fig. 1. We choose the base to be a $b\times c$
rectangular region and $a$ to be the plate separation (the distance
from the base to the piston). The piston divides the volume into two
regions. We refer to region I as the interior and region II as the
exterior. Both regions contribute to the Casimir force on the
piston. The Casimir piston therefore modifies some previous standard
Casimir results \cite{Bordagreport} where the effects of the
exterior region are not included.

The Casimir piston for the electromagnetic field with
perfect-conductor conditions in a three-dimensional rectangular
cavity (box) was studied recently \cite{Kardar} and it was shown
that the Casimir force on the piston is again attractive (in
contrast to results without exterior region where the force could be
positive). The piston for perfect-conductor conditions including the
effects of temperature was studied further in \cite{Marachevsky1,
Marachevsky2, Marachevsky3} where among other things,  the long and
short distance behavior of the free energy was investigated. A
theorem was obtained in \cite{Klich}, where it was shown that the
Casimir force between two bodies related by reflection is always
attractive, independent of the exact form of the bodies or
dielectric properties. This theorem was then generalized further in
\cite{Bachas} where it was shown that reflection positivity implies
that the force between any mirror pair of charge-conjugate probes of
the quantum vacuum is attractive. Attraction does not occur in all
Casimir piston scenarios. In a recent paper \cite{Barton}, the
Casimir piston for a weakly reflecting dielectric was considered and
it was shown that though attraction occurred for small plate
separation, this could switch to repulsion for sufficiently large
separation. Moreover, for thick enough material, the force remained
attractive for all plate separations in agreement with the results
in \cite{Kardar}. Two recent preprints \cite{Fulling,Zhai} also
discuss scenarios where repulsive Casimir forces in pistons can be
achieved.

For the case of a massless scalar field in a three-dimensional
cavity, approximate expressions for the Casimir force were obtained
valid for small plate separation \cite{Kardar}. In this paper, we
consider the general case of arbitrary lengths. We present exact
expressions for the Casimir force on a piston due to a massless
scalar field obeying Dirichlet boundary conditions in a
three-dimensional box with sides of arbitrary lengths $a,b$ and $c$.
We find that the Casimir force on the piston is negative and runs
from $-\infty$ (in the limit $a\!\to\!0$) to $0$ (in the limit
$a\!\to\!\infty$). For small plate separation $a$, we recover the
results found in \cite{Kardar}. We also obtain an exact alternative
expression for the Casimir force that is useful computationally when
the plate separation is large. We focus our attention on Dirichlet
instead of Neumann boundary conditions because it is the more
interesting case of the two. It is clear that Neumann boundary
conditions will yield a negative Casimir force since the
contribution from both the interior and exterior are negative. It is
not a priori obvious that in the Dirichlet case the Casimir force
will be negative because there exists values of the ratios $a/c$ and
$b/c$ where the interior contributes a positive (repulsive) Casimir
force. It is therefore interesting to see that in such cases the
exterior contributes a negative force of larger magnitude with the
important consequence that the total Casimir force is negative. It
is worth mentioning that the study of massless scalar fields is not
only of theoretical interest but has direct relevance to physical
systems such as Bose-Einstein condensates
\cite{Ariel2,Pomeau,Visser}.

The Casimir energy can be viewed as the energy with boundary
conditions (a sum over discrete modes) minus the energy without
boundary conditions (a volume integral over continuous modes). The
sum over the discrete modes can typically be decomposed into a
volume divergent term (the continuum part that can be subtracted), a
surface divergent term and a finite part. In previous set-ups
without region II, finite results were obtained by throwing out the
surface divergent term. Though the finite results agreed with the
zeta function regularization technique, there is nothing that can
physically justify throwing out the surface term. It yields a
cut-off dependent Casimir force that cannot be removed via a
renormalization of the physical parameters of the theory
\cite{Jaffe, Barton2,Barton3}. The agreement between zeta function
regularization and cut-off technique (with surface term thrown out)
occurs because the zeta function technique in effect renormalizes
the surface term to zero. The Casimir piston resolves this issue
satisfactorily by having the exterior and interior contributions to
the surface divergence cancel. This has been demonstrated in
Refs.\cite{Cavalcanti,Kardar} and we assume this cancellation to
hold here. One can simply calculate the Casimir force $F_1$ and
$F_2$ on the piston due to region I and II respectively without
including the cut-off dependent terms. The total Casimir force on
the piston can then be obtained by adding $F_1$ and $F_2$. One must
just keep in mind that $F_1$ and $F_2$ actually have cut-off
dependent terms but that they cancel when the two are added.

There are two positive aspects to the Casimir piston: the exterior
is now included in the calculation of the Casimir force (we add
$F_2$ to $F_1$) and the surface divergence is handled via a
cancellation procedure instead of simply throwing it out.

We work in units where $\hbar=c=1$ ($c$ is the speed of light). Note
that from now on, when the variable $c$ appears in the text, it
always refers to one of the lengths of the base (see Fig. 1).

\section{Casimir piston in three dimensions: exact results}
\setcounter{equation}{0}

The Casimir energy $E_D$ for massless scalar fields in a
$d$-dimensional box of arbitrary lengths $L_1,...,L_d$ obeying
Dirichlet boundary conditions can be conveniently expressed as an
analytical part -- composed of Riemann zeta and gamma functions --
plus a sum of over Bessel functions (eq. \reff{ED22}\,; see appendix
A and Refs.\cite{Ariel,Wolfram}): \beq
 E_D =
 \dfrac{\pi}{2^{d+1}}\sum_{j=0}^{d-1} (-1 )^{d+j}\,\,\xi^{\,d-1}_{\,k_1,.., k_j}\,\Big\{\dfrac{L_{k_1}\ldots L_{k_j}}{(L _d)^{j+1}}\Big(
\Gamma(\tfrac{j+2}{2})\,\pi^{\frac{-j-4}{2}}\, \zeta(j+2) + R_j
\Big)\,\Big\}\eeq{ED2} where $R_j$ represents the sum over modified
Bessel functions $K_{\nu}$:  \beq R_j
=\sum_{n=1}^{\infty}\,\sumprime_{\substack{l_i=-\infty\\i=1,\ldots,
j}}^{\infty}\dfrac{2\,\,n^{\frac{j+1}{2}}}{\pi}\,\dfrac{\,K_{\frac{j+1}{2}}
\big(\,2\pi\,n\,\sqrt{(\ell_1\frac{L_{k_1}}{L_d})^2+\cdots+(\ell_j\,\frac{L_{k_j}}{L_d})^2}\,\,\,\big)}
{\left[(\ell_1\frac{L_{k_1}}{L_d})^2+\cdots+(\ell_j\frac{L_{k_j}}{L_d})^2\right]^{\tfrac{j+1}{4}}}\,\,.
\eeq{rjbb} The prime in the sum for $R_j$ means that the case where
all $\ell$'s are simultaneously zero is excluded. Note that $R_j$ is
a function of the ratios of the lengths. In \reff{ED2}, there is an
implicit summation over the integers $k_i$.  The symbol
$\xi^{\,d-1}_{\,k_1,.., k_j}$ is defined as \beq
\xi^{\,d-1}_{\,k_1,.., k_j}=\begin{cases} 1&\word{if}
 k_1 \!<\!k_2\!<\! \ldots\! <k_j \,;\,1 \le k_j \le d-1\\ 0&
\word{otherwise}.
\end{cases}\eeq{ordered}
The above symbol apparently does not have a name and we refer to it
as the $ordered$ symbol in Appendix A. The ordered symbol ensures
that the implicit sum over the $k_i$ in \reff{ED2} is over all
distinct sets $\{k_1,\ldots, k_j\}$, where the $k_i$ are integers
that can run from $1$ to $d-1$ inclusively under the constraint that
$k_1\!<\!k_2\!<\!\cdots\!<k_j$. The superscript $d\!-\!1$ specifies
the maximum value of $k_j$. For example, if $j=2$ and $d=4$ then
$\xi^{\,d-1}_{\,k_1,.., k_j}=\xi^{\,3}_{\,k_1, k_2}$ and the
non-zero terms are $\xi_{\,1,2}$ , $\xi_{\,1,3}$ and $ \xi_{\,2,3}$.
This means the summation is over $\{k_1,k_2\}=(1,2), (1,3)$ and
$(2,3)$. Note that the implicit summation over $k_i$ is also
performed in $R_j$ since $R_j=R_j(L_{k_1}/L_d,..,L_{k_j}/L_d)$. For
the special case of $j=0$, $R_j$ is defined to be zero and
$\xi^{\,d-1}_{\,k_1,.., k_j}$ and $L_{k_j}$ are defined to be
identically one so that $\xi^{\,d-1}_{\,k_1,..,
k_j}\,\frac{L_{k_1}\ldots L_{k_j}}{(L _d)^{j+1}}$ = $1/L_d$ for
$j=0$.

From \reff{ED2} we can readily obtain the Dirichlet Casimir energy
in three dimensions ($d=3$): \beq
E_{D}=-\dfrac{\pi^2}{1440}\,\dfrac{L_1\,L_2}{L_3^3}+\dfrac{\zeta(3)}{32\,\pi\,L_3^2}\big(L_1
+L_2\big)-\dfrac{\pi}{96\,L_3} +R(L_1,L_2,L_3) \eeq{dirich} where
$R$ is a function of $L_1,L_2$ and $L_3$ and represents the sums
over $R_j$'s i.e.\beq
R(L_1,L_2,L_3)=\dfrac{\pi}{16\,L_3^2}\Big[L_1\, R_1(L_1/L_3)+
L_2\,R_1(L_2/L_3)\,\Big]
-\dfrac{\pi\,L_1\,L_2}{16\,L_3^3}\,R_2(L_1/L_3,L_2/L_3)\eeq{RR}
where $R_1(L_1/L_3)$ means that $R_1$ is a function of $L_1/L_3$.
The functions $R_1$ and $R_2$ are sums over modified Bessel
functions given by \reff{rjbb} i.e. \beq\begin{split}
&R_1(L_1/L_3)=\sum_{n=1}^{\infty}\sum_{\ell=1}^{\infty}\,\dfrac{4\,n}{\pi\,\ell}\,\frac{L_3}{L_1}\,K_1\Big(2\,\pi\,n\,\ell\,\frac{L_1}{L_3}\Big)\\&
R_2(L_1/L_3,L_2/L_3)=\sum_{n=1}^{\infty}\,\sumprime_{\ell_1,\ell_2=-\infty}^{\infty}\!\!
\dfrac{2\,n^{3/2}\,K_{3/2}\Big(2\,\pi\,n\,\sqrt{\Big(\ell_1\dfrac{L_1}{L_3}\Big)^2+\Big(\ell_2\dfrac{L_2}{L_3}\Big)^2}\,\,\Big)}
{\pi\left[\Big(\ell_1\dfrac{L_1}{L_3}\Big)^2+\Big(\ell_2\dfrac{L_2}{L_3}\Big)^2\right]^{3/4}}\,.
\end{split}\eeq{RRR}
The Casimir energy does not depend on which sides are labeled $L_1$,
$L_2$ and $L_3$. Expression \reff{dirich} for the Casimir energy is
therefore invariant under permutations of the labels $L_1$, $L_2$
and $L_3$ and we are free to label the three sides as we wish. For
the Casimir piston depicted in Fig. 1, there are two regions to
consider. In region I, the three sides are $a$, $b$ and $c$ and we
label them $L_1=c$, $L_2=b$ and $L_3=a$. In region II, the three
sides are $s-a$, $c$ and $b$ and we label them $L_1=s-a$, $L_2=c$
and $L_3=b$. The Dirichlet Casimir energy in region I and II is
obtained by substituting the corresponding lengths in \reff{dirich}:
\beq\begin{split}
E_{D1}&=-\dfrac{\pi^2}{1440}\,\dfrac{b\,c}{a^3}+\dfrac{\zeta(3)}{32\,\pi\,a^2}(b+c)-\dfrac{\pi}{96\,a}
+R(c,b,a)\\
E_{D2}&=-\dfrac{\pi^2}{1440}\,\dfrac{(s-a)\,c}{b^3}+\dfrac{\zeta(3)}{32\,\pi\,b^2}(s-a
+c)-\dfrac{\pi}{96\,b} +R(s-a,c,b)\,.\end{split} \eeq{dirich1} The
function  $R(c,b,a)$ is obtained from \reff{RR} and \reff{RRR}: \beq
\begin{split}R(c,b,a)&=\dfrac{\pi}{16\,a^2}\big[c\, R_1(c/a)+ b\,R_1(b/a)\,\big]
-\dfrac{\pi\,c\,b}{16\,a^3}\,R_2(c/a,b/a)\\
&\quad=\dfrac{1}{4\,a}\sum_{n=1}^{\infty}\sum_{\ell=1}^{\infty}\,\dfrac{n}{\ell}\,\Big[
K_1\big(2\,\pi\,n\,\ell\,c/a\big)+K_1\big(2\,\pi\,n\,\ell\,b/a\big)\Big]\\
&\qquad\qquad-\dfrac{\,b\,c}{8\,a^3}\,\sum_{n=1}^{\infty}\,\sumprime_{\ell_1,\ell_2=-\infty}^{\infty}\!\!
\dfrac{n^{3/2}\,K_{3/2}\Big(2\,\pi\,n\,\sqrt{\Big(\dfrac{\ell_1\,c}{a}\Big)^2+
\Big(\dfrac{\ell_2\,b}{a}\Big)^2}\,\,\Big)}
{\left[\Big(\dfrac{\ell_1\,c}{a}\Big)^2+\Big(\dfrac{\ell_2\,b}{a}\Big)^2\right]^{3/4}}\,
\end{split}\eeq{Rcba}
where the prime in the sum means that the case
$\ell_1\!=\!\ell_2\!=\!0$ is excluded from the sum.\footnote{Only
the case when $\ell_1$ and $\ell_2$ are simultaneously zero is to be
excluded. In particular, one can have $\ell_1=0$ when $\ell_2\ne0$
and vice versa.} The Casimir force on the piston is obtained by
taking the derivative with respect to the plate separation $a$ and
then taking the limit $s\to \infty$\,:
\beq\begin{split} F&=-\frac{\partial}{\partial\,a}(E_{D1}+E_{D2})\\
&=-\frac{\pi^2\,b\,c}{480\,a^4}\!+\frac{\zeta(3)\,(b+
c)}{16\,\pi\,a^3} \!-\frac{\pi}{96\,a^2}\!-\!
\frac{\pi^2\,c}{1440\,b^3}+\frac{\zeta(3)}{32\,\pi\,b^2}
\!-\!R'(c,b,a)\!-\!\lim_{s\to\infty}\!\!R'(s\!-\!a,c,b)\,.
\end{split}\eeq{force}
We now evaluate the last two terms in \reff{force}.
$R'(c,b,a)\equiv\frac{\partial}{\partial\,a}\,R(c,b,a)$ can readily
be obtained by taking the derivative of \reff{Rcba} with respect to
$a$: \beq
\begin{split}&R'(c,\!b,\!a)\!=\!\dfrac{1}{4\,a}\sum_{n=1}^{\infty}\!\sum_{\ell=1}^{\infty}\dfrac{n}{\ell}\Big[
K_1'\big(2\pi n\ell c/a\big)\!+\!K_1'\big(2\pi n \ell
b/a\big)\!-\!\dfrac{1}{a}K_1\big(2\pi n \ell c/a\big)
\!-\!\dfrac{1}{a}K_1\big(2\pi n \ell b/a\big)\Big]\\
&\!+\!\dfrac{b\,c}{8\,a^{3/2}}\sum_{n=1}^{\infty}\sumprime_{\ell_1,\ell_2=-\infty}^{\infty}\!\!
\Big\{\dfrac{3\,n^{3/2}\,K_{3/2}\Big(\dfrac{2\,\pi\,n}{a}\,\sqrt{\ell_1^2\,c^2
\!+\!
\ell_2^2\,b^2}\,\Big)}{2\,a\,\left(\,\ell_1^{\,2}\,c^2\!+\!\ell_2^{\,2}\,b^2\,\right)^{3/4}}-
\dfrac{n^{3/2}\,K_{3/2}'\Big(\dfrac{2\,\pi\,n}{a}\,\sqrt{\ell_1^2\,c^2
\!+\!
\ell_2^2\,b^2}\,\Big)}{\left(\,\ell_1^{\,2}\,c^2\!+\!\ell_2^{\,2}\,b^2\,\right)^{3/4}}\Big\}\,
\end{split}\eeq{Rcba_prime}
where a prime on the Bessel functions denotes derivative with
respect to the plate separation $a$. The last term in \reff{force}
can be written as \beq\begin{split}
&\lim_{s\to\infty}\!R'(s\!-\!a,c,b)\equiv
\lim_{s\to\infty}\frac{\partial}{\partial\,a}\,R(s-a,c,b)=
-\lim_{u\to\infty}\frac{\partial}{\partial\,u}\,R(u,c,b)\\&=
-\lim_{u\to\infty}\frac{\partial}{\partial\,u}\,
\Bigg\{\dfrac{1}{4\,b}\sum_{n=1}^{\infty}\sum_{\ell=1}^{\infty}\,\dfrac{n}{\ell}\,\Big[
K_1\big(2\,\pi\,n\,\ell\,u/b\big)+K_1\big(2\,\pi\,n\,\ell\,c/b\big)\Big]\\
&\qquad\qquad\qquad-\dfrac{\,u\,c}{8\,b^3}\,\sum_{n=1}^{\infty}\,\sumprime_{\ell_1,\ell_2=-\infty}^{\infty}\!\!
\dfrac{n^{3/2}\,K_{3/2}\Big(2\,\pi\,n\,\sqrt{\Big(\dfrac{\ell_1\,u}{b}\Big)^2+
\Big(\dfrac{\ell_2\,c}{b}\Big)^2}\,\,\Big)}
{\left[\Big(\dfrac{\ell_1\,u}{b}\Big)^2+\Big(\dfrac{\ell_2\,c}{b}\Big)^2\right]^{3/4}}\,\,\Bigg\}
\end{split}\eeq{Rtr}
where the substitution $u=s\!-\!a$ was made and $R(u,c,b)$ was
obtained from \reff{Rcba} by substituting the appropriate lengths.
The modified Bessel functions and their derivatives decrease
exponentially fast so that the only term in \reff{Rtr} that survives
is the case $\ell_1=0$ in the double sum. With $\ell_1=0$, the
remaining sum over $\ell_2$ does not include zero and can be
replaced by twice the sum from $1$ to $\!\infty$. One therefore
obtains \beq
\lim_{s\to\infty}\!R'(s\!-\!a,c,b)=\dfrac{\,c}{4\,b^3}\,\sum_{n=1}^{\infty}\,\sum_{\ell=1}^{\infty}
\Big(\dfrac{n\,b}{\ell\,c}\Big)^{3/2}K_{3/2}(2\,\pi\,n\,\ell\,c/b)\,\,.
\eeq{remainBC}

After substituting \reff{remainBC} into \reff{force}, the Casimir
force on the piston is  \beq\begin{split} F&=
-\frac{\pi^2\,b\,c}{480\,a^4}+\frac{\zeta(3)\,(b+ c)}{16\,\pi\,a^3}
-\frac{\pi}{96\,a^2}-R'(c,b,a)\\& \qquad\qquad-
\frac{\pi^2\,c}{1440\,b^3}+\frac{\zeta(3)}{32\,\pi\,b^2}
-\dfrac{\,c}{4\,b^3}\,\sum_{n=1}^{\infty}\,\sum_{\ell=1}^{\infty}
\Big(\dfrac{n\,b}{\ell\,c}\Big)^{3/2}K_{3/2}(2\,\pi\,n\,\ell\,c/b)\,\,.
\end{split}\eeq{force2}
Eq. \reff{force2} is an exact expression for the Casimir force on
the piston for Dirichlet boundary conditions. No approximations have
been made. With $R'(c,b,a)$ given by \reff{Rcba_prime}, one can
calculate exactly the force for any values of $a$, $b$ and $c$. Note
that the second row in \reff{force2} has no dependence on $a$ and
corresponds to the contribution from region II. If we set $b\!=\!c$
and take the small $a$ limit ($a\!\!<\!<\!\!b$), we recover the
expression for the Casimir force obtained in \cite{Kardar}. In this
limit $R'(c,b,a)$ is exponentially suppressed (exactly zero in the
limit $a\to 0$) and with  $b=c$, the second row in \reff{force2}
yields $0.004831546/c^2$ in agreement with Dirichlet results in
\cite{Kardar}.

When $a$ is sufficiently large, $R'(c,b,a)$ dominates over the other
$a$-dependent terms in \reff{force2}. In fact, in the limit $a \!\to
\!\infty$, the other $a$-dependent terms vanish while $R'(c,b,a)$
reduces to a finite function of $b$ and $c$. Therefore, a full
analysis of the Casimir force on the piston -- one that goes beyond
small values of $a$ -- requires one to have the exact expression
\reff{Rcba_prime} for $R'(c,b,a)$.

In \reff{force2}, the first and second row are the contributions
from region I and II respectively: \beq\begin{split}
F_1&=-\dfrac{\pi^2\,b\,c}{480\,a^4}+\dfrac{\zeta(3)\,(b+
c)}{16\,\pi\,a^3} -\dfrac{\pi}{96\,a^2}-R'(c,b,a)
\\\word{and}\, F_{2}&=-
\dfrac{\pi^2}{1440}\,\dfrac{c}{b^3}+\dfrac{\zeta(3)}{32\,\pi\,b^2}-
\dfrac{\,c}{4\,b^3}\,\sum_{n=1}^{\infty}\,\sum_{\ell=1}^{\infty}
\Big(\dfrac{n\,b}{\ell\,c}\Big)^{3/2}K_{3/2}(2\,\pi\,n\,\ell\,c/b)\,.\end{split}
\eeq{F1F2}

To compute $F_1$ and $F_2$ we specify the two ratios $a/c$ and $b/c$
and express results in units of $1/c^2$. Let us look at the case of
the cube: $a/c=1$ and $b/c=1$. Using  \reff{Rcba_prime}, we obtain
$R'(c,b,a)=-0.000214214$. The last term in $F_2$ -- the sum over the
Bessel function -- yields $-0.000271643$. The remaining analytical
terms in $F_1$ and $F_2$ can easily be evaluated. $F_1$ and $F_2$
for the case of the cube is given by \beq\begin{split}
F_{1_{cube}}&= -.005458275 + 0.000214214= -0.005244061\\
F_{2_{cube}}&=0.005103189-0.000271643=0.004831546\,.\end{split}\eeq{Fcube}
We see that the Casimir force from region I is attractive and the
force from region II is repulsive. Clearly, region II weakens
significantly the total Casimir force. However, $F_2$ is not large
enough to reverse the sign and the Casimir force remains attractive:
\beq F_{cube}= F_1+F_2= -0.000412515\,.\eeq{fcube2}

The force $F_1$ can actually  be positive (repulsive) \cite{Cheng}.
For example, if $a/c=0.1$ and $b/c=0.1$ then $F_1=+3.80553076$.
However, the force due to the second region is then negative and
larger in magnitude: $F_2=-5.65818384$. Adding the contribution from
region II therefore causes a reversal of sign to take place. Though
$F_1$ is positive, the total Casimir force, $F=F_1+F_2$, is negative
and equal to $-1.85265308$.

The expression for the Casimir force on the piston,
eq.\reff{force2}, is valid for any positive values of $a$, $b$ and
$c$ but is most useful computationally when the plate separation $a$
is the smallest of the three lengths. The ratios $b/a$ and $c/a$ are
then greater than or equal to one (we are also free to label the
sides of the base such that $c\ge b$ so that $c/b$ is also greater
than or equal to one). The sums over the Bessel functions and their
derivatives in $\reff{Rcba_prime}$ then converge exponentially fast
yielding accurate and quick results. In Appendix B we derive an
alternative expression $F_{alt}$ for the Casimir force on the piston
that is useful computationally when the plate separation $a$ is not
the smallest of the three lengths. The alternative expression is
given by \reff{alter}: \beq\begin{split}
F_{alt}&=-\dfrac{1}{4\,b}\sum_{n=1}^{\infty}\sum_{\ell=1}^{\infty}\,\dfrac{n}{\ell}\,K_1'\big(2\,\pi\,n\,\ell\,a/b\big)\\
&+\dfrac{\partial}{\partial\,a}\,\Bigg[\dfrac{a\,c}{4\,b^3}\,\sum_{n=1}^{\infty}\,\sum_{\ell_1=1}^{\infty}\sum_{\ell_2=-\infty}^{\infty}\!\!
\dfrac{n^{3/2}\,K_{3/2}\Big(2\,\pi\,n\,\sqrt{\Big(\dfrac{\ell_1\,a}{b}\Big)^2+\Big(\dfrac{\ell_2\,c}{b}\Big)^2}\,\,\Big)}
{\left[\Big(\dfrac{\ell_1\,a}{b}\Big)^2+\Big(\dfrac{\ell_2\,c}{b}\Big)^2\right]^{3/4}}\Bigg]\,
\end{split}\eeq{alter2}
where the prime above the modified Bessel function $K_1$ implies
partial derivative with respect to $a$:
$K_1'(2\,\pi\,n\,\ell\,a/b)\equiv
\frac{\partial}{\partial\,a}\,K_1(2\,\pi\,n\,\ell\,a/b)$. As before,
we are free to label the base such that $c\ge b$. If the plate
separation $a$ is not the smallest length, it follows that $a\ge b$
and the above sums over Bessel functions and their derivatives
converge exponentially fast. Both expressions, \reff{force2} and
\reff{alter2}, yield the same value for the Casimir force. However,
computationally, expression \reff{force2} is better to use if $a$ is
the smallest length and the alternative expression \reff{alter2} is
better to use otherwise.

For a given value of $b$ and $c$, the Casimir force $F$ on the
piston ranges from $-\infty$ to zero corresponding to the two
extreme limits of the plate separation $a$ i.e. \beq \lim_{a\to0} F
=-\infty\quad \word{and}\quad \lim_{a\to\infty} F=0\,. \eeq{limits}
The first limit in \reff{limits} follows readily if one uses
expression \reff{force2} for the Casimir force. In the limit $a
\!\to\! 0$, the $-1/a^4$ term dominates and goes to $-\infty$ (note
that $\lim_{a\to0} R'(c,b,a)=0$). The second limit in \reff{limits}
follows readily if one uses the alternative expression $F_{alt}$
given by \reff{alter2}. In the limit $a\!\to\! \infty$, $F_{alt}$ is
clearly zero since the Bessel functions and their derivatives
decrease exponentially fast to zero as already mentioned at the end
of Appendix B. One can also understand this latter result
intuitively: as $a\to \infty$, region I becomes equivalent to region
II and the forces from each region balance each other out i.e. $\lim
_{a\to\infty}F_1=-F_2$.

A plot of the Casimir force $F$ versus $a/c$ is shown in Fig. 2 for
the case $b/c=1$ (the force is in units of $1/c^2$). The Casimir
force is negative, has a large magnitude at small values of $a/c$
and decreases rapidly in magnitude towards zero as $a/c$ increases
in agreement with the two limits given by \reff{limits}. One obtains
a similar plot for any value of $b/c$. A 3D plot of $F$ versus $a/c$
and $b/c$ is shown in Fig. 3. The Casimir force is negative
throughout and a slice taken at any value of $b/c$ yields a similar
profile to the 2D plot in Fig.2 with the magnitude of the force
shifting to greater values as $b/c$ increases. For any given slice,
the Casimir force lies between the two limits given by
\reff{limits}.

\begin{appendix}
\section{Explicit expression for Dirichlet Casimir energy for $d$-dimensional box
with sides of arbitrary lengths}
\def\theequation{A.\arabic{equation}}
\setcounter{equation}{0}

In this appendix, we derive the explicit formula \reff{ED2} for the
Casimir energy of massless scalar fields confined to a
$d$-dimensional box of arbitrary lengths for Dirichlet boundary
conditions. We begin by stating explicit formulas for the
$d$-dimensional Casimir energy obeying periodic boundary conditions
\cite{Ariel}. The second step is to express the Dirichlet energy as
a sum over the periodic energy \cite{Ariel,Wolfram} \footnote
{\cite{Ariel} uses a multidimensional cut-off technique and
\cite{Wolfram} uses the Epstein zeta function \cite{Epstein}
technique. This technique has been developed extensively over the
years \cite{Wolfram}-\cite{Neto} and there are some excellent books
on the subject \cite{book1, book2, book3}.}. The third step, the
main part of the appendix,  is to perform explicitly this sum to
obtain the compact expression \reff{ED22} for the Dirichlet energy.

The Casimir energy for massless scalar fields in a $d$-dimensional
box of arbitrary lengths $L_1,...,L_d$ and periodic boundary
conditions can be explicitly expressed as an analytical part --
composed of Riemann zeta and gamma functions -- plus a sum of over
Bessel functions \cite{Ariel}: \beq\begin{split}&E_{p_{ _{\,L_1,..,
L_d}}}\,(d) =-\,\pi\sum_{j=0}^{d-1}\dfrac{L_1\ldots L_j}{(L
_{j+1})^{j+1}}\Big( \Gamma(\tfrac{j+2}{2})\,\pi^{\frac{-j-4}{2}}\,
\zeta(j+2) + R_j \Big)\\&=
\,\dfrac{-\pi}{6\,L_1}-\dfrac{\zeta(3)}{2\,\pi}\dfrac{L_1}{L_2^2}-\dfrac{\pi^2}{90}\dfrac{L_1\,L_2}{L_3^3}+
\cdots-R_1\,\dfrac{\pi\,L_1}{L_2^2}
-R_2\,\dfrac{\pi\,L_1\,L_2}{L_3^3}+\cdots
\end{split}\eeq{epfinal} where $R_j$ represents the sum over modified Bessel functions
$K_{\nu}$: \beq R_j
=\sum_{n=1}^{\infty}\,\sumprime_{\substack{l_i=-\infty\\i=1,\ldots,
j}}^{\infty}\dfrac{2\,\,n^{\frac{j+1}{2}}}{\pi}\,\dfrac{\,K_{\frac{j+1}{2}}
\big(\,2\pi\,n\,\sqrt{(\ell_1\frac{L_1}{L_{j+1}})^2+\cdots+(\ell_j\,\frac{L_j}{L_{j+1}})^2}\,\,\,\big)}
{\left[(\ell_1\frac{L_1}{L_{j+1}})^2+\cdots+(\ell_j\frac{L_j}{L_{j+1}})^2\right]^{\tfrac{j+1}{4}}}\,\,.
\eeq{rjbb2} The prime in the above sum means that the case where all
$\ell$'s are zero is excluded. Note that for $j\!=\!0$ one sets
$R_j$ to zero and $L_j$ identically to one so that $\frac{L_1\ldots
L_j}{(L _{j+1})^{j+1}}$ is equal to $1/L_1$ for $j=0$. Note also
that $R_j$ is a function of ratios of lengths i.e.
$R_j=R_j(L_1/L_{j+1},...,L_j/L_{j+1})$. The notation $E_{p_{
_{\,L_1,\ldots, L_d}}}\,(d)$ is a compact way of saying that the
Casimir energy $E_p$ is a function of the dimension $d$ and the
lengths $L_1,...,L_d$.

Our goal is to obtain a similar explicit expression for the case of
Dirichlet boundary conditions. We begin by noting that the Dirichlet
case can be expressed as a sum over the periodic Casimir energies
$E_p$ (see \cite{Ariel,Wolfram}): \beq E_D =
\dfrac{1}{2^{d+1}}\sum_{m=1}^d \,(-1 )^{d+m}\!\!\!\!\!\!
\sum_{\substack{\{k_1,..,k_m\}\\k_1<k_2<\cdots<k_m\\\ k_m\le
d}}E_{p_{_{\,L_{k_1},..,L_{k_m}}}}(m)\,. \eeq{Edirich}

The sum over the $k_i$'s is over all sets $\{k_1,.., k_m\}$, where
the $k_i$ are integers that can run from $1$ to a maximum value of
$d$ under the constraint that $k_1\!<\!k_2\!<\!\cdots\!<k_m$. To
specify that $d$ is the maximum value we write $k_m\le d$ under the
sum in \reff{Edirich}. $E_{p_{_{L_{k_1},..,L_{k_m}}}}(m)$ is the
periodic energy \reff{epfinal} replacing $d$ by $m$ and $L_1$ by
$L_{k_1}$, $L_2$ by $L_{k_2}$, etc. Note that the replacement $L_1$
by $L_{k_1}$, etc. must also be performed inside $R_j$ given by
\reff{rjbb2}. The above notation for the sum over $k_i$ is
cumbersome. It is convenient to introduce a symbol
$\xi^{\,d}_{\,k_1,.., k_m}$ defined by \beq \xi^{\,d}_{\,k_1,..,
k_m}=\begin{cases} 1&\word{if}
 k_1 \less k_2 \less \ldots\! <k_m \,;\,1 \le k_i \le d\\ 0&
\word{otherwise}.
\end{cases}\eeq{final}
The superscript $d$ specifies the dimension which is the maximum
value of $k_m$. The above defined symbol apparently does not have a
name and for simplicity we shall refer to it as the $ordered$
symbol. Equation \reff{Edirich} can now be conveniently expressed
with the $ordered$ symbol:
 \beq E_D =
\dfrac{1}{2^{d+1}}\sum_{m=1}^d \,(-1 )^{d\!+\!m}
\,\xi^{\,d}_{\,k_1,.., k_m}\,E_{p_{_{\,L_{k_1},..,L_{k_m}}}}(m)\,
\eeq{Edirich2} where $implicit$ $summation$ over the $k_i$'s is
assumed. After substituting \reff{epfinal} into \reff{Edirich2} one
obtains \beq E_D = \dfrac{-\pi}{2^{d+1}}\sum_{m=1}^d \,(-1 )^{d+m}
\xi^{\,d}_{\,k_1,.., k_m} \,\Big\{\sum_{j=0}^{m-1}
\dfrac{L_{k_1}\ldots L_{k_j}}{(L _{k_{j+1}})^{j+1}}\Big(
\Gamma(\tfrac{j+2}{2})\,\pi^{\frac{-j-4}{2}}\, \zeta(j+2) + R_j
\Big) \Big\}\eeq{Edirich3} where $R_j$ is the function \reff{rjbb2}
with $L_1$ replaced by $L_{k_1}$, $L_2$ by $L_{k_2}$, etc. For
simplicity we define \beq f_{j_{_{\,\,k_1,..,k_{j+1}}}} \equiv
\dfrac{L_{k_1}\ldots L_{k_j}}{(L _{k_{j+1}})^{j+1}}\Big(
\Gamma(\tfrac{j+2}{2})\,\pi^{\frac{-j-4}{2}}\, \zeta(j+2) + R_j
\Big)\eeq{fk1} and rewrite \reff{Edirich3} as  \beq
E_D=\dfrac{-\pi}{2^{d+1}}\sum_{m=1}^d \sum_{j=0}^{m-1}(-1 )^{d+m}\,
\xi^{\,d}_{\,k_1,.., k_m} f_{j_{_{\,\,k_1,..,k_{j+1}}}}\!\!\!\!
=\dfrac{\pi}{2^{d+1}}\sum_{j=0}^{d-1} \sum_{m=j+1}^{d}(-1
)^{d+m+1}\,\xi^{\,d}_{\,k_1,.., k_m} f_{j_{_{\,\,k_1,..,k_{j+1}}}}
\eeq{EDD} where we have rewritten the limits on each sum. We can
decompose $\xi^{\,d}_{\,k_1,.., k_m}$ into a sum of two terms:
$\xi^{\,d-1}_{\,k_1,.., k_{m-1},d}+\xi^{\,d-1}_{\,k_1,.., k_m}$. The
first term, $\xi^{\,d-1}_{\,k_1,.., k_{m-1},d}\,\,$, means that
$k_m$ is set to its maximum value of $d$ and the sum is now over the
remaining $k_i$'s with $k_{m-1}$ having a maximum possible value of
$d-1$. With $k_m=d$ in the first term, the maximum possible value of
$k_m$ in the second term is $d-1$ (hence the superscript $d-1$ in
the second term). Note that for the special case $m=d$, the above
decomposition yields only one term not two terms i.e.
$\xi^{\,d}_{\,k_1,.., k_d}=\xi^{\,d-1}_{\,k_1,.., k_{d-1},d} + 0$
since $k_d$ can only be equal to $d$.

With this decomposition the sum over $m$ becomes \beq
\begin{split}&\sum_{m=j+1}^{d}(-1 )^{d+m+1}\,\,
\xi^{\,d}_{\,k_1,.., k_m} = \sum_{m=j+1}^{d}(-1 )^{d+m+1}\,\,
\big[\,\xi^{\,d-1}_{\,k_1,.., k_{m-1},d} + \xi^{\,d-1}_{\,k_1,..,
k_{m}}\,\big]
\\&= (-1)^{d+j}\big[\,\xi^{\,d-1}_{\,k_1,.., k_j,d}+
(\,\xi^{\,d-1}_{\,k_1,.., k_{j+1}}-\xi^{\,d-1}_{\,k_1,..,
k_{j+1},d}\,)-(\,\xi^{\,d-1}_{\,k_1,..,
k_{j+2}}-\xi^{\,d-1}_{\,k_1,..,
k_{j+2},d}\,)\\&\quad\quad+\ldots(-1)^{d-j}(\,\xi^{\,d-1}_{\,k_1,..,
k_{d-1}}-\xi^{\,d-1}_{\,k_1,.., k_{d-1},d})\,\big]\,.
\end{split}\eeq{oops}
The two terms inside each pair of round brackets in \reff{oops} have
opposite signs and cancel each other \footnote{In the first pair of
round brackets $\xi^{\,d-1}_{\,k_1,.., k_{j+1}}$ cancels with
$-\xi^{\,d-1}_{\,k_1,.., k_{j+1},d}\,$. The fact that $k_{j+2}$ is
equal to $d$ in the latter term is irrelevant since the summation
over $f_{j_{_{\,\,k_1,..,k_{j+1}}}}$ in \reff{EDD} stops at
$k_{j+1}$ for a given $j$. Therefore $\xi^{\,d-1}_{\,k_1,..,
k_{j+1},d}$ is equivalent to $\xi^{\,d-1}_{\,k_1,.., k_{j+1}}$. The
same logic applies to the terms inside the other round brackets.} so
that the sum over $m$ reduces to only the first term \beq
\sum_{m=j+1}^{d}(-1 )^{d+m+1}\,\, \xi^{\,d}_{\,k_1,.., k_m}
=(-1)^{d+j}\,\xi^{\,d-1}_{\,k_1,.., k_j,d}\,.\eeq{xiD} The Dirichlet
Casimir energy is obtained by substituting \reff{xiD} in \reff{EDD}:
\beq E_D=\dfrac{\pi}{2^{d+1}}\!\sum_{j=0}^{d-1}(-1
)^{d+j}\,\xi^{\,d-1}_{\,k_1,.., k_j,d}\,
f_{j_{_{\,\,k_1,..,k_{j+1}}}} \!\!\!\!=\!
\dfrac{\pi}{2^{d+1}}\sum_{j=0}^{d-1} (-1
)^{d+j}\,\xi^{\,d-1}_{\,k_1,.., k_j}\,
f_{j_{_{\,\,k_1,\,..,\,k_j,\,d}}}\,. \eeq{fD} The function
$f_{j_{_{\,\,k_1,\,..,\,k_j,\,d}}}$ is obtained by setting $k_{j+1}$
equal to $d$ in \reff{fk1}. We finally obtain our explicit
expression for the Dirichlet Casimir energy \beq
 E_D =
 \dfrac{\pi}{2^{d+1}}\sum_{j=0}^{d-1} (-1 )^{d+j}\,\,\xi^{\,d-1}_{\,k_1,.., k_j}\,\Big\{\dfrac{L_{k_1}\ldots L_{k_j}}{(L _d)^{j+1}}\Big(
\Gamma(\tfrac{j+2}{2})\,\pi^{\frac{-j-4}{2}}\, \zeta(j+2) + R_j
\Big)\,\Big\}\eeq{ED22} where $R_j$ is given by \reff{rjbb2} with
$L_1\to L_{k_1}$, $L_{j+1}\to L_{k_{j+1}}=L_d$ i.e. \beq R_j
=\sum_{n=1}^{\infty}\,\sumprime_{\substack{l_i=-\infty\\i=1,\ldots,
j}}^{\infty}\dfrac{2\,\,n^{\frac{j+1}{2}}}{\pi}\,\dfrac{\,K_{\frac{j+1}{2}}
\big(\,2\pi\,n\,\sqrt{(\ell_1\frac{L_{k_1}}{L_{d}})^2+\cdots+(\ell_j\,\frac{L_{k_j}}{L_{d}})^2}\,\,\,\big)}
{\left[(\ell_1\frac{L_{k_1}}{L_{d}})^2+\cdots+(\ell_j\frac{L_{k_j}}{L_{d}})^2\right]^{\tfrac{j+1}{4}}}\,\,.
\eeq{rjbb3} For the case $j=0$, $R_j$ is zero and
$\xi^{\,d-1}_{\,k_1,.., k_j}$ and $L_{k_j}$ are defined as unity.

\section{Alternative expression for Casimir force on piston}
\def\theequation{B.\arabic{equation}}
\setcounter{equation}{0}

One can derive an alternative expression for the Casimir force $F$
on the piston by labeling the lengths $L_1$, $L_2$ and $L_3$
differently. We are free to label the lengths in any way we want
since the Casimir energy is invariant under permutations of $L_1$,
$L_2$ and $L_3$. In region I, the three lengths are $a$, $b$ and $c$
and we label them now $L_1=a$, $L_2=c$ and $L_3=b$. In region II,
the three lengths are $s-a$, $b$ and $c$ and we label them now
$L_1=s-a$, $L_2=c$ and $L_3=b$. The Dirichlet Casimir energy in
region I and II is then obtained via \reff{dirich} \beq\begin{split}
E_{D1}&=-\dfrac{\pi^2}{1440}\,\dfrac{a\,c}{b^3}+\dfrac{\zeta(3)}{32\,\pi\,b^2}(a+c)-\dfrac{\pi}{96\,b}
+R(a,c,b)\\
E_{D2}&=-\dfrac{\pi^2}{1440}\,\dfrac{(s-a)\,c}{b^3}+\dfrac{\zeta(3)}{32\,\pi\,b^2}(s-a
+c)-\dfrac{\pi}{96\,b} + R(s-a,c,b)\end{split} \eeq{dirich5} where
$R(a,c,b)$ and $R(s-a,c,b)$ are defined via \reff{RR} and
\reff{RRR}. The Casimir force $F_1$ due to region I and $F_2$ due to
region II (with $s\to \infty$) are \beq\begin{split}
F_1&=-\dfrac{\partial}{\partial\,a}E_{D1}
=\dfrac{\pi^2}{1440}\,\dfrac{c}{b^3}-\dfrac{\zeta(3)}{32\,\pi\,b^2}-R'(a,c,b)\\
F_2&=\lim_{s\to\infty}-\dfrac{\partial}{\partial\,a}E_{D2}=-\dfrac{\pi^2}{1440}\,\dfrac{c}{b^3}+\dfrac{\zeta(3)}{32\,\pi\,b^2}
-\lim_{s\to\infty}R'(s-a,c,b)\,.
\end{split}\eeq{FRR12}
The total Casimir force $F=F_1+F_2$ on the piston is then simply
\beq F=-R'(a,c,b)-\lim_{s\to\infty}R'(s-a,c,b) \eeq{FRR} where the
prime denotes derivative with respect to the plate separation $a$.
Note that the analytical terms -- the Riemann zeta and gamma terms
-- have canceled. This is also what occurs in the two-dimensional
Casimir piston (see \cite{Cavalcanti}). The second term in
\reff{FRR} has already been obtained and is given by
\reff{remainBC}. The function $R(a,c,b)$ can be obtained from the
function $R(c,b,a)$ given by \reff{Rcba} by replacing  $c$ with $a$,
$b$ with $c$ and $a$ with $b$ i.e. \beq
\begin{split}R(a,c,b)&=\dfrac{1}{4\,b}\sum_{n=1}^{\infty}\sum_{\ell=1}^{\infty}\,\dfrac{n}{\ell}\,\Big[
K_1\big(2\,\pi\,n\,\ell\,a/b\big)+K_1\big(2\,\pi\,n\,\ell\,c/b\big)\Big]\\
&\qquad\qquad-\dfrac{\,a\,c}{8\,b^3}\,\sum_{n=1}^{\infty}\,\sumprime_{\ell_1,\ell_2=-\infty}^{\infty}\!\!
\dfrac{n^{3/2}\,K_{3/2}\Big(2\,\pi\,n\,\sqrt{\Big(\dfrac{\ell_1\,a}{b}\Big)^2+\Big(\dfrac{\ell_2\,c}{b}\Big)^2}\,\,\Big)}
{\left[\Big(\dfrac{\ell_1\,a}{b}\Big)^2+\Big(\dfrac{\ell_2\,c}{b}\Big)^2\right]^{3/4}}\,.
\end{split}\eeq{bts2}
The derivative of $R(a,c,b)$ with respect to the plate separation
$a$
 is \beq\begin{split}
R'(a,c,b)&= \dfrac{1}{4\,b}\sum_{n=1}^{\infty}\sum_{\ell=1}^{\infty}\,\dfrac{n}{\ell}\,K_1'\big(2\,\pi\,n\,\ell\,a/b\big)\\
&-\dfrac{\partial}{\partial\,a}\,\Bigg[\dfrac{a\,c}{8\,b^3}\,\sum_{n=1}^{\infty}\,\sumprime_{\ell_1,\ell_2=-\infty}^{\infty}\!\!
\dfrac{n^{3/2}\,K_{3/2}\Big(2\,\pi\,n\,\sqrt{\Big(\dfrac{\ell_1\,a}{b}\Big)^2+\Big(\dfrac{\ell_2\,c}{b}\Big)^2}\,\,\Big)}
{\left[\Big(\dfrac{\ell_1\,a}{b}\Big)^2+\Big(\dfrac{\ell_2\,c}{b}\Big)^2\right]^{3/4}}\Bigg]\,.
\end{split}\eeq{diffR2}
With  $R'(a,c,b)$ given by \reff{diffR2} and
$\lim_{s\to\infty}R'(s-a,c,b)$ given by \reff{remainBC}, the Casimir
force \reff{FRR} yields
\beq\begin{split} F&=-\dfrac{1}{4\,b}\sum_{n=1}^{\infty}\sum_{\ell=1}^{\infty}\,\dfrac{n}{\ell}\,K_1'\big(2\,\pi\,n\,\ell\,a/b\big)\\
&+\dfrac{\partial}{\partial\,a}\,\Bigg[\dfrac{a\,c}{8\,b^3}\,\sum_{n=1}^{\infty}\,\sumprime_{\ell_1,\ell_2=-\infty}^{\infty}\!\!
\dfrac{n^{3/2}\,K_{3/2}\Big(2\,\pi\,n\,\sqrt{\Big(\dfrac{\ell_1\,a}{b}\Big)^2+\Big(\dfrac{\ell_2\,c}{b}\Big)^2}\,\,\Big)}
{\left[\Big(\dfrac{\ell_1\,a}{b}\Big)^2+\Big(\dfrac{\ell_2\,c}{b}\Big)^2\right]^{3/4}}\Bigg]\\
&-\dfrac{c}{4\,b^3}\,\sum_{n=1}^{\infty}\,\sum_{\ell_2=1}^{\infty}
\dfrac{n^{3/2}\,K_{3/2}\Big(2\,\pi\,n\,\ell_2\,c/b \Big)}
{\Big(\dfrac{\ell_2\,c}{b}\Big)^{3/2}}\,.\end{split} \eeq{FG} The
above expression has three terms and it can be simplified by
noticing that the $\ell_1=0$ case in the second term cancels out
with the last term. The Casimir force on the piston reduces to the
following final expression: \beq\begin{split}
F_{alt}&=-\dfrac{1}{4\,b}\sum_{n=1}^{\infty}\sum_{\ell=1}^{\infty}\,\dfrac{n}{\ell}\,K_1'\big(2\,\pi\,n\,\ell\,a/b\big)\\
&+\dfrac{\partial}{\partial\,a}\,\Bigg[\dfrac{a\,c}{4\,b^3}\,\sum_{n=1}^{\infty}\,\sum_{\ell_1=1}^{\infty}\sum_{\ell_2=-\infty}^{\infty}\!\!
\dfrac{n^{3/2}\,K_{3/2}\Big(2\,\pi\,n\,\sqrt{\Big(\dfrac{\ell_1\,a}{b}\Big)^2+\Big(\dfrac{\ell_2\,c}{b}\Big)^2}\,\,\Big)}
{\left[\Big(\dfrac{\ell_1\,a}{b}\Big)^2+\Big(\dfrac{\ell_2\,c}{b}\Big)^2\right]^{3/4}}\Bigg]\,.
\end{split}\eeq{alter}
The above is our alternative expression for the Casimir force on the
piston. It is valid for any positive values of $a$, $b$ and $c$ but
it is especially useful computationally when $a$ is not the smallest
of the three lengths. We are free to label the base such that $c\ge
b$. If $a$ is not the smallest length, then the ratios $a/b$ and
$c/b$ are both greater than or equal to one. This ensures that the
sums over the Bessel functions and their derivatives in \reff{alter}
will converge exponentially fast making computations easy and
accurate. Note that the sum over $\ell_1$ and the sum over $\ell$ in
\reff{alter} do not include zero. Therefore as $a$ increases the
Bessel functions and their derivatives will always decrease
exponentially and reach zero in the limit $a\to \infty$. The Casimir
force on the piston is therefore zero in the limit $a\to \infty$.
\end{appendix}
%%%%%%%%%%%%%%%%%%%%%%%%%%%%%%%%%%%%%%%%%%%%%%%%%%%%%%%%%
\section*{Acknowledgments}

I thank the Natural Sciences and Engineering Research Council of
Canada (NSERC) for their financial support.

%%%%%%%%%%%%%%%%%%%%%%%%%%%%%%%%%%%%%%%%%%%%%%%%

\clearpage

\begin{figure}[ht]
\begin{center}
\includegraphics[scale=1]{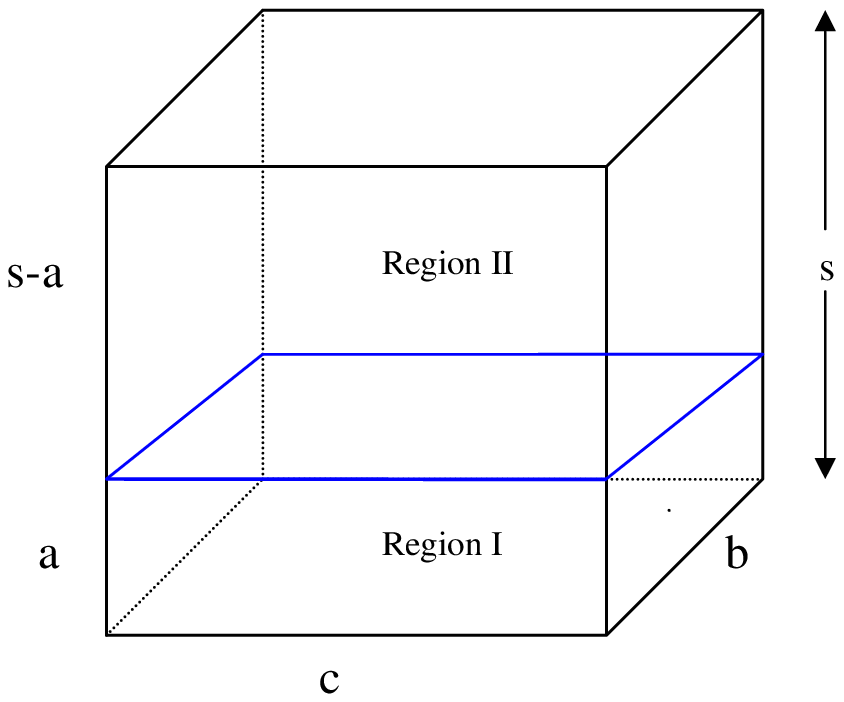}
\caption{Casimir piston in three dimensions.}
\end{center}
\end{figure}

\begin{figure}[ht]
\begin{center}
\includegraphics[scale=0.5]{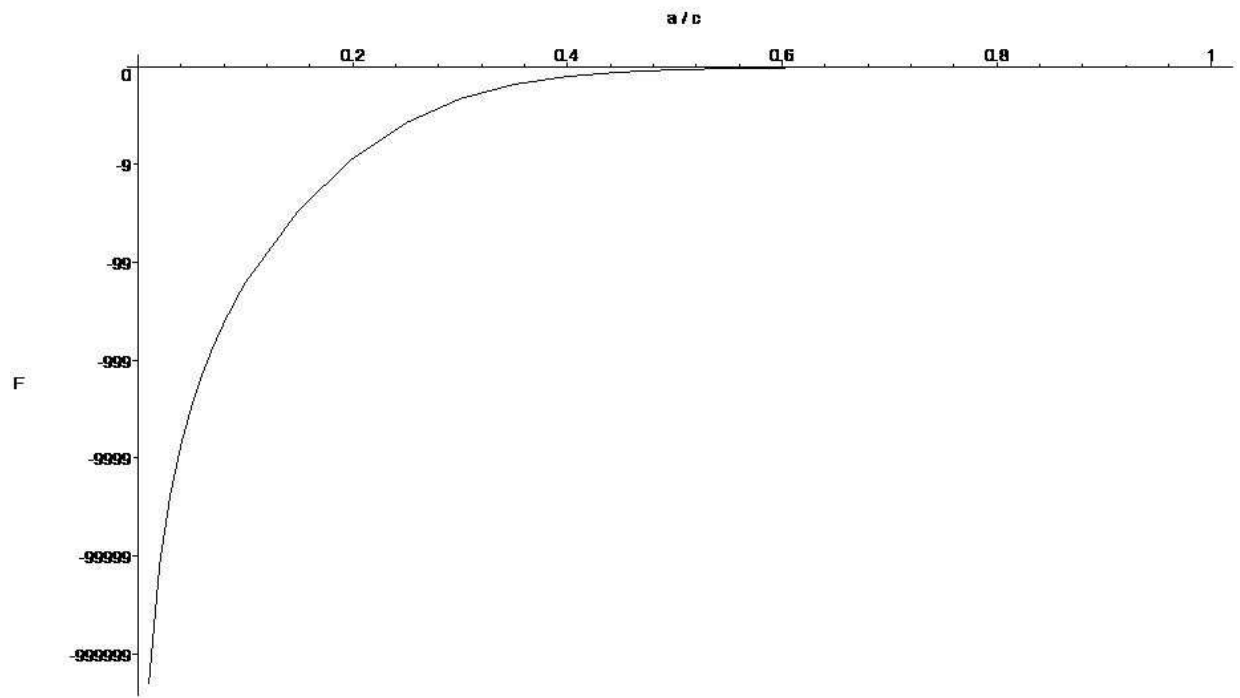}
\caption{Casimir force $F$ versus $a/c$ for the case $b/c=1$ where
$a$ is the plate separation and $b$ and $c$ are the sides of the
base. The force is in units of $1/c^2$. The force is large and
negative at small values of $a/c$ and remains negative with its
magnitude decreasing quickly to zero as $a/c$ increases. One obtains
a similar plot for any value of $b/c$ (see the 3D plot below)}
\end{center}
\end{figure}

\begin{figure}[ht]
\begin{center}
\includegraphics[scale=0.5]{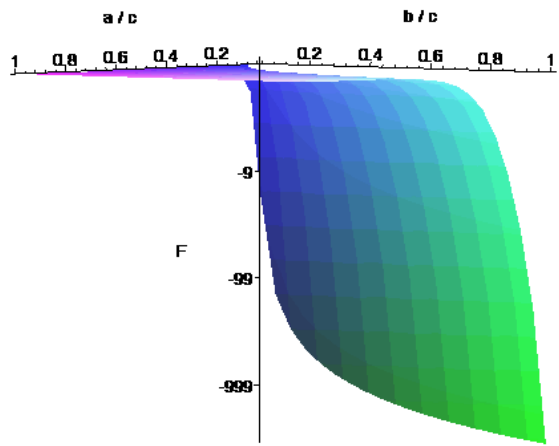}
\caption{3D plot of Casimir force $F$ versus $a/c$ and $b/c$. The
force is in units of $1/c^2$. For a given $b/c$, the profile is the
same as in the 2D plot: the force is large and negative at small
values of $a/c$ and remains negative with its magnitude decreasing
quickly to zero as $a/c$ increases. The value of $b/c$ shifts the
magnitude of the force towards larger values as it increases.}
\end{center}
\end{figure}

\clearpage
\end{document}